# Emergence of multiple quasi-ferromagnetic magnon modes induced by strong magnetoelastic coupling in $TmFeO_3$ single crystal

Sourabh Manna[1#], Felix Fuhrmann[1#], Sergio Rodríguez Fernández[1], Xiaoxuan Ma[2,3,*], Haiyang Chen[4], Luca M. Carrella[5], , Edgar Galindez-Ruales[1], Gerhard Jakob[1], Olena Gomonay[1], Shixun Cao[2,6,*] and Mathias Kläui[1,7,*]

[1]Institute of Physics, Johannes Gutenberg University Mainz, Staudingerweg 7, Mainz, 55128, Germany.

[2]Materials Genome Institute, Shanghai University, Shanghai 200444, China.

[3]School of Materials Science and Engineering, Shanghai University, Shanghai 200444, China.

[4]Key Laboratory of Artificial Structures and Quantum Control (Ministry of Education), School of Physics and Astronomy, Shanghai Jiao Tong University, Shanghai 200240, China.

[5]Department of Chemistry, Pharmacy and Geosciences, Johannes Gutenberg Univ Mainz, Duesbergweg 10-14, D-55128 Mainz, Germany.

[6]Shanghai Key Laboratory of High Temperature Superconductors, Shanghai University, Shanghai 200444, China.

[7]Center for Quantum Spintronics, Norwegian University of Science and Technology, Trondheim & 7491, Norway.

# These authors contributed equally.

Corresponding authors' email: maxiaoxuan@shu.edu.cn, sxcao@shu.edu.cn, klaeui@uni-mainz.de

## Abstract

We investigate the magnetization dynamics of $TmFeO_3$ single crystals across the spin-reorientation phase transition using broadband microwave absorption spectroscopy up to 87.5 GHz. Temperature- and magnetic-field-dependent antiferromagnetic resonance measurements reveal the characteristic softening of the quasi-ferromagnetic (q-FM) resonance mode at the $\Gamma_2 \rightarrow \Gamma_{24}$ and $\Gamma_{24} \rightarrow \Gamma_4$ transition points. The finite magnon gap observed at the transition points reflects the strong magnetoelastic coupling. In addition to the uniform q-FM mode, multiple magnon modes appear in the intermediate $\Gamma_{24}$ phase, separated by approximately 0.5–2 GHz and exhibiting similar field and temperature dependence. These additional modes are attributed to nonuniform spin-wave excitations arising from the periodic magnetic domain structure present in the intermediate phase and their hybridization with acoustic phonons mediated by strong magnetoelastic coupling. Our results demonstrate that the spin-reorientation transition in $TmFeO_3$ provides a natural platform for generating multiple hybridized magnon modes, offering new opportunities for tunable magnonic excitations in rare-earth orthoferrites.

**R**are-earth orthoferrites $R\text{FeO}_3$ where $R$ represents the rare-earth ion ($R^{3+}$), have attracted continued interest owing to their diverse magnetic properties including weak ferromagnetism arising from the Dzyaloshinskii-Moria interaction [1,2], multiferroicity [3,4], pronounced magneto-optical effects [5,6], strong magnetoelastic coupling [7–10] and ultrafast spin dynamics [11–15]. Recently, these materials have also gained renewed attention because their crystal symmetry leads to d-wave altermagnetism, opening a new paradigm of magnetic order and the associated spin transport phenomena in orthoferrite systems [16–19]. A distinctive magnetic characteristic of the rare-earth orthoferrites is their tendency to switch among three symmetry-allowed magnetic configurations of the Fe subsystem–denoted as $\Gamma_2$, $\Gamma_{24}$ and $\Gamma_4$. Transitions between these states occurs through a continuous second order spin-reorientation phase transition (SRPT) driven by the magnetic interactions between $Fe^{3+}$ and $R^{3+}$ spin subsystems [20–22]. In the vicinity of the SRPT, the strong interplay among magnetic, acoustic, and optical degrees of freedom leads to interesting phenomena such as emergence of coupled magnetoelastic wave and THz electric-field-driven ballistic spin switching [20,23,24].

Thulium orthoferrite ($TmFeO_3$) is one of the rare-earth orthoferrites exhibiting strong magnetoelastic coupling and long-range antiferromagnetic order up to 635 K [25]. In this material, the magnetoelastic interaction strongly modifies the magnetization dynamics in the SRPT region, making magnetic resonance an effective probe of the underlying spin-lattice coupling. Previous studies using THz pulse excitation have demonstrated the SRPT and the associated softening of magnetic resonance modes near the phase transition points [11,12,26]. However, intense THz pulses can induce local heating, which limits precise temperature control of the sample–an important requirement for studying the SRPT region. Microwave absorption techniques provide a significant advantage in this regard because they allow for low-power excitation of magnetic resonance modes while maintaining uniform sample temperature. Moreover, these studies did not discuss the effect of any external magnetic field although it is crucial for tunability of magnetization dynamics. Earlier studies have reported GHz and sub THz microwave-excitation of magnetic resonance in $TmFeO_3$, but they are largely restricted to either low frequencies or limited magnetic-field ranges, leaving the combined influence of temperature and magnetic field on the resonance modes insufficiently explored [27–29].

Understanding the magnetoelastic modulation of the magnetization dynamics in the SRPT region therefore requires systematic investigation of magnetic resonance as a function of both temperature and external magnetic field, particularly focusing on the quasi-ferromagnetic resonance mode, which is strongly influenced by spin–lattice coupling. In this Letter, we report a comprehensive study of the temperature- and magnetic-field-dependent quasi-ferromagnetic resonance modes of $TmFeO_3$ up to 87.5 GHz across the SRPT region using low-power microwave excitation. Our results reveal that strong magnetoelastic coupling in the intermediate phase of the spin-reorientation transition leads to the emergence of multiple quasi-ferromagnetic magnon modes, providing new insight into spin–lattice hybridization in rare-earth orthoferrites.

$TmFeO_3$ is a collinear antiferromagnet whose crystallographic symmetry belongs to the orthorhombic space group $Pnma$ (No. 62) [30]. The magnetic structure of $TmFeO_3$ can be described by two antiferromagnetically coupled sublattices with magnetizations $\boldsymbol{M}_1$ and $\boldsymbol{M}_2$ corresponding to the magnetic moments localized on the Fe ions. Equivalently, the magnetic order can be expressed in terms of the Néel vector $\boldsymbol{n} = \boldsymbol{M}_1 - \boldsymbol{M}_2$ which characterizes the antiferromagnetic order and the net magnetization $\boldsymbol{m} = \boldsymbol{M}_1 + \boldsymbol{M}_2$ arising from the weak canting of the spins. The magnetic field-temperature ($H$-$T$) phase diagram of $TmFeO_3$ includes three magnetic phases as illustrated in Fig. 1a. At low temperature ($T < T_1 \approx 85$ K) [31,32] the system adopts the $\Gamma_2$ phase with $\boldsymbol{n} \parallel c$-axis and $\boldsymbol{m} \parallel a$-axis. At high temperature ($T > T_2 \approx 93$ K) [31,32], the materials stabilizes in the $\Gamma_4$ phase with $\boldsymbol{n} \parallel a$-axis and $\boldsymbol{m} \parallel c$-axis. An intermediate $\Gamma_{24}$ phase exists at temperature $T_1 < T < T_2$ where $\boldsymbol{n}$ and $\boldsymbol{m}$ continuously rotate in the $ac$-plane [31,32]. This spin reorientation phase transition can be induced by varying the temperature ($\Gamma_2 \rightarrow \Gamma_{24} \rightarrow \Gamma_4$) or by applying an external magnetic field along $a$ or $c$-axes [28]. For instance, an external magnetic field applied along the $c$-axis at $T < T_1$ can induce the ($\Gamma_2 \rightarrow \Gamma_{24} \rightarrow \Gamma_4$) phase transition. Conversely, the ($\Gamma_4 \rightarrow \Gamma_{24} \rightarrow \Gamma_2$) phase transition can be driven by applying the external magnetic field along $a$-axis at $T > T_2$. For a given temperature and a magnetic field $\boldsymbol{H}_{DC}$, the equilibrium magnetic state can be obtained by minimizing the total magnetic free energy given by [28]:

$$F = \frac{1}{2M_s} H_{ex} \boldsymbol{m}^2 + \frac{1}{2M_s} H_c n_c^2 + \frac{1}{2M_s^3} H_4 n_c^4 + \frac{H_{d1}}{M_s} n_c m_a - \frac{H_{d2}}{M_s} n_a m_c - \boldsymbol{H}_{DC} \cdot \boldsymbol{m} \quad (1)$$

Here $n_a(n_c)$ and $m_a(m_c)$ are the components of the $\boldsymbol{n}$ and $\boldsymbol{m}$ along the $a(c)$-axis respectively. $H_{ex}$ is the intersublattice exchange field that keeps $\boldsymbol{M}_1$ and $\boldsymbol{M}_2$ antiparallel and $M_s = 2|\boldsymbol{M}_1|$. The parameters $H_c$ and $H_4$ describe the magnetic anisotropy imposed by the orthorhombic symmetry. $H_{d1}$ and $H_{d2}$ parametrize the DMI interactions and the last term in Eq. 1 represents the Zeeman energy. Note that a smooth reorientation of the Néel vector (and the magnetization) occurs in the spin-flop configuration when $\boldsymbol{H}_{DC}$ is applied parallel to the initial orientation of the Néel vector. Such behavior arises from the relationship between the DMI coefficients: $H_{d1} \approx H_{d2} \gg |H_{d1} - H_{d2}|$.

In magnetic equilibrium, magnons can be excited via the coupling of the magnetic moments with a microwave excitation field ($\boldsymbol{h}_{rf}$). In the long wavelength limit, the magnon spectra of $TmFeO_3$ consists of two distinct branches: (i) the low frequency quasi-ferromagnetic (q-FM) mode corresponds to the circularly polarized precession of $\boldsymbol{m}$ ($\boldsymbol{m} \perp \boldsymbol{h}_{rf}$) accompanied by the oscillation of the Néel vector and (ii) the high frequency quasi antiferromagnetic (q-AFM) mode characterized by the linearly polarized oscillation of $\boldsymbol{m}$ ($\boldsymbol{m} \parallel \boldsymbol{h}_{rf}$) together with the precession of the Néel vector. The resonant mode ($\boldsymbol{k} = 0$) of the q-FM branch shows significant frequency variation across the SRPT and softens at the $\Gamma_2 \rightarrow \Gamma_{24}$ and $\Gamma_{24} \rightarrow \Gamma_4$ transition points [21,26,28]. In contrast, the frequency of the q-AFM resonant mode hardly changes across the ($\Gamma_2 \rightarrow \Gamma_{24} \rightarrow \Gamma_4$) phase transition [21,26]. $TmFeO_3$ is also known for its strong magnetoelastic coupling which induces pronounced spontaneous strains and softening of the elastic moduli and the associated phonon velocity in the vicinity of the phase transition region [33]. This magnetoelastic coupling substantially affects the softening of the q-FM magnons leading to finite magnon gaps at the $\Gamma_2 \rightarrow \Gamma_{24}$ and $\Gamma_{24} \rightarrow \Gamma_4$ transition points [8,24,34]. Hence, the influence of magnetoelastic coupling on the q-FM resonant modes is particularly significant in the intermediate $\Gamma_{24}$ phase across the SRPT. To investigate these effects, we performed microwave-absorption-based ferromagnetic resonance measurements on a (001)-oriented single crystal of $TmFeO_3$.

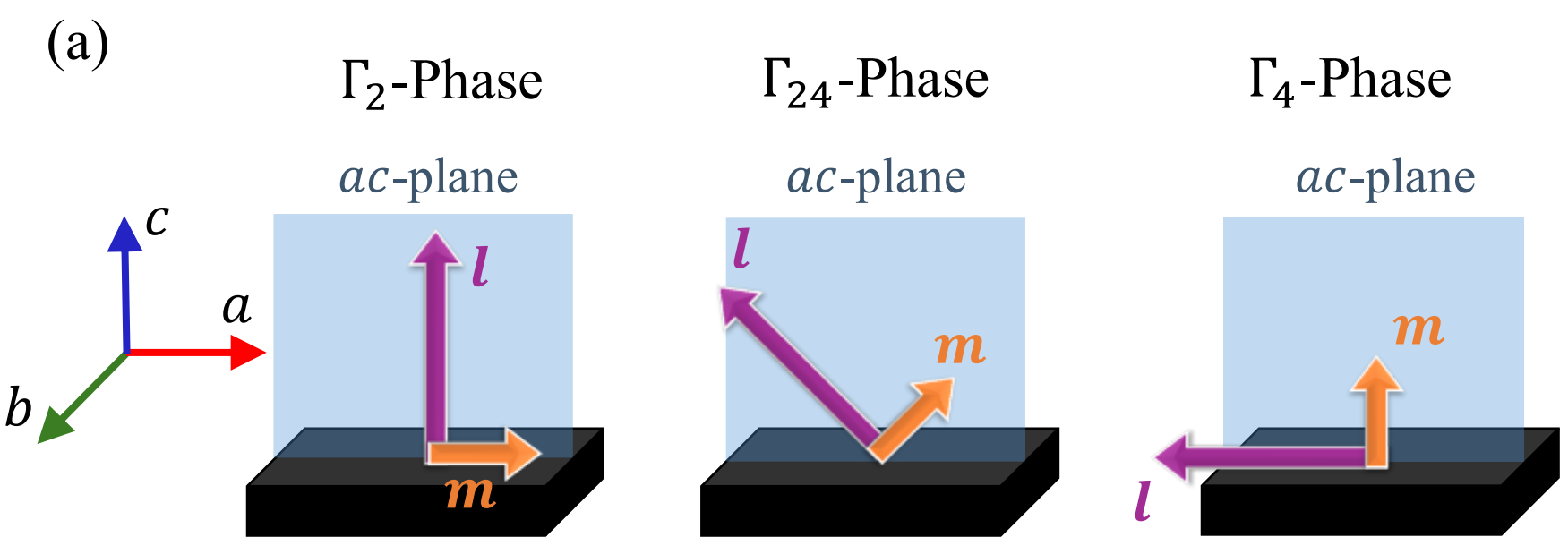

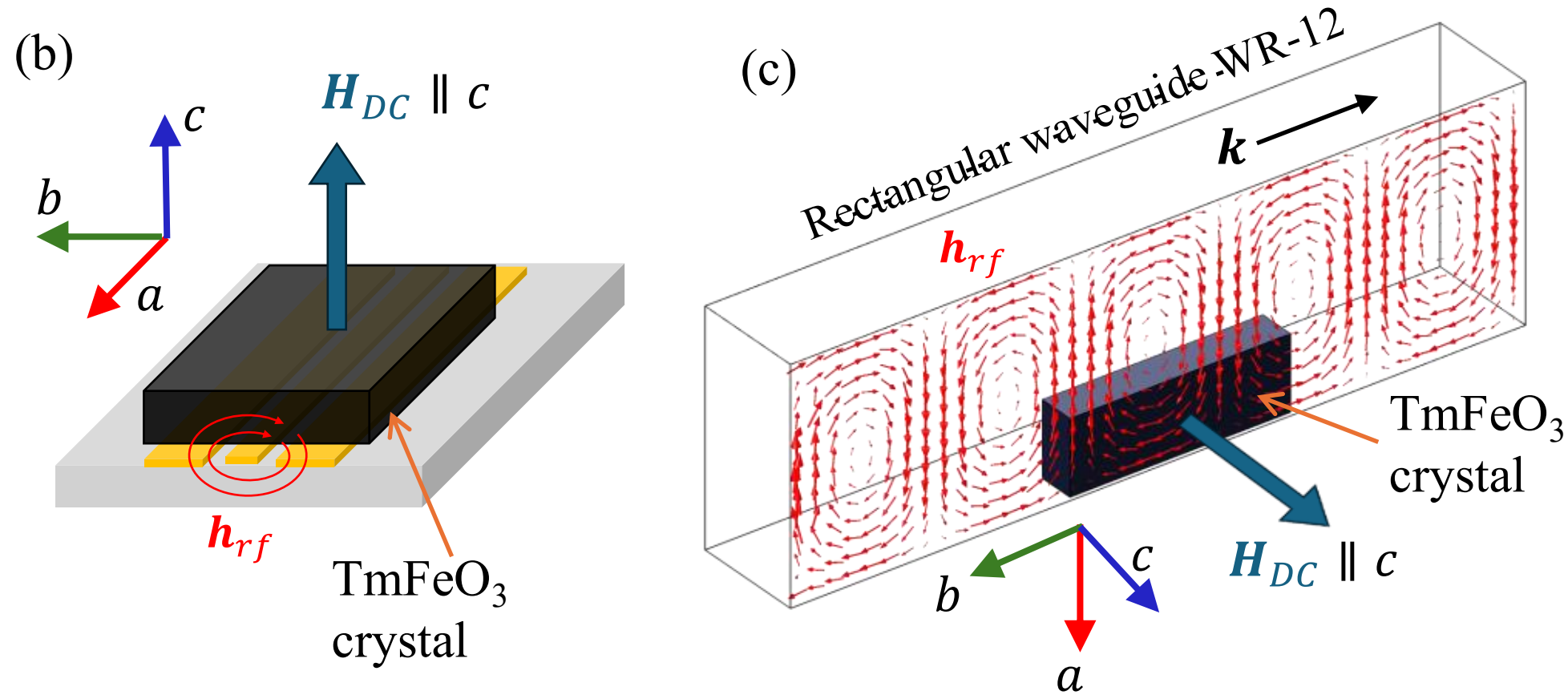

**Figure 1** (a) Spin reorientation phase transition ($\Gamma_2 \rightarrow \Gamma_{24} \rightarrow \Gamma_4$) in $TmFeO_3$ illustrated by the orientation of the Néel vector ($\boldsymbol{n}$) and the weak moment ($\boldsymbol{m}$) (b) Orientation of the microwave excitation field $\boldsymbol{h}_{rf}$ and the external magnetic field $\boldsymbol{H}_{DC}$ for (a) flip-chip ferromagnetic resonance measurement up to 40 GHz and (c) rectangular waveguide-based ferromagnetic resonance measurement between 70.5 GHz to 87.5 GHz.

Single crystal $TmFeO_3$ was grown by the optical floating-zone technique from polycrystalline $TmFeO_3$ synthesized by a conventional solid-state reaction between high-purity $Tm_2O_3$ (99.99%) and $Fe_2O_3$ (99.99%) powders. Details of the crystal growth and sample preparation are discussed in the supplementary information. The crystal was cut along the (001)-direction ($c$-axis) and was polished subsequently. To investigate the magnetization dynamics in the vicinity of SRPT, variable temperature microwave transmission-based antiferromagnetic resonance (AFMR) measurements were performed over a broad frequency range (up to 87.5 GHz) using two complementary experimental configurations.

For measurements up to 40 GHz, AFMR spectra were obtained using the flip-chip ferromagnetic resonance (FMR) technique using a coplanar waveguide (CPW) [35,36]. The $TmFeO_3$ single crystal was mounted on a CPW sample holder inside a cryostat equipped with a superconducting electromagnet capable of generating a uniaxial dc magnetic field up to 8 T. The sample was placed directly on the CPW such that it symmetrically covered the central signal line, ensuring efficient coupling between the microwave magnetic field and the magnetic moments. The two terminals of the CPW were connected to the two ports of a Rhode & Schwarz ZVA40 vector network analyzer (VNA).

In this flip-chip configuration, the dc magnetic field ($\boldsymbol{H}_{DC}$) was applied along the crystallographic $c$-axis (out of plane), while the microwave magnetic field was oriented along the $b$-axis, as determined by the CPW geometry (see Fig. 1b). The sample temperature was controlled using the variable temperature insert (VTI) of the cryostat using an Oxford Instruments ITC 503 temperature controller. An additional calibrated Lakeshore Cryotronics temperature sensor attached to the sample holder ensured accurate temperature determination. After stabilizing both the VTI and the sample temperatures, the transmission parameter $S_{21}$was recorded as a function of frequency at fixed magnetic fields. By sweeping $\boldsymbol{H}_{DC}$, two-dimensional microwave absorption spectra were obtained as a function of frequency and magnetic field. The raw $S_{21}$ data were subsequently processed using the derivative-divide method [36] to enhance the visibility of the resonance modes and obtain clear AFMR spectra.

To access the high-frequency regime between 70 and 90 GHz, microwave absorption measurements were performed using a WR-12 rectangular waveguide-based setup. The $TmFeO_3$ single crystal sample was placed inside a WR-12 waveguide (see Fig. 1c) integrated into a custom-designed sample holder compatible with the cryostat environment. In this geometry, the magnetic moments in the sample were excited by the microwave magnetic field associated with the fundamental $TE_{10}$ mode propagating through the waveguide. The dc magnetic field was again applied along the crystallographic $c$-axis, while the microwave magnetic field was within the ab-plane, as illustrated schematically in Fig.1c. Temperature control was achieved using the same VTI-based approach as mentioned before, with an additional Lakeshore temperature sensor attached to the sample holder. However, the exact temperature of the sample was not determined as no temperature sensor could be placed on the sample itself inside the waveguide. High-frequency microwave transmission measurements were carried out using the VNA in combination with custom-built frequency extenders and a frequency synthesizer acting as the local oscillator for harmonic mixing. With the local oscillator frequency fixed at 10 GHz, the setup provided access to the 70.5–75.5 GHz and 81.5–87.5 GHz frequency bands. For each temperature, the system was allowed to stabilize before recording the transmission parameter $S_{21}$ as a function of magnetic field and frequency. The resulting spectra were processed using the derivative-divide method to enhance the visibility of the AFMR modes.

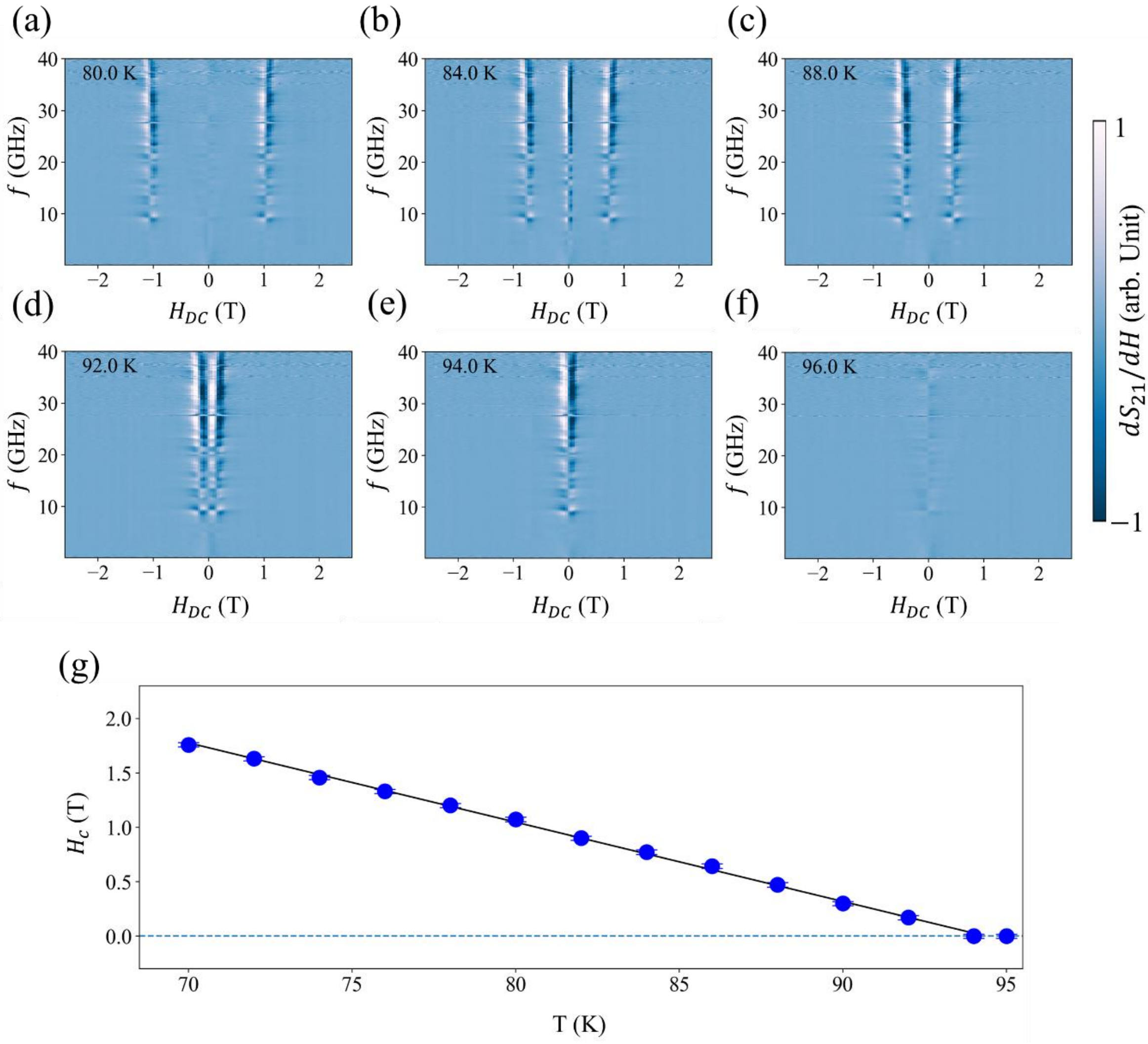

**Figure. 2**. (a) – (f) Resonance spectra of $TmFeO_3$ up to 40 GHz for $H_{DC} \parallel c$-axis as obtained from the flip-chip ferromagnetic resonance experiment. (g) Variation of critical field $H_c$ as a function of temperature.

In our experimental geometry, the microwave magnetic field $\boldsymbol{h}_{rf}$ lies within the $ab$-plane of the $TmFeO_3$ crystal while $\boldsymbol{H}_{DC} \parallel c$-axis (see Fig. 1b and c). Since the net magnetic moment $\boldsymbol{m}$ undergoes reorientation in the $ac$-plane, the $\boldsymbol{h}_{rf}$-component which is orthogonal to $\boldsymbol{m}$ can couple to the low-frequency q-FM resonant (q-FMR) mode [28,34,37]. This mode appears well below 100 GHz around the $\Gamma_2 \rightarrow \Gamma_{24}$ and $\Gamma_{24} \rightarrow \Gamma_4$ phase transition points [11,26] and therefore, accessible in our experimental frequency range. Figure 2(a)–(f) depicts the microwave absorption spectra obtained from the temperature dependent flip-chip AFMR measurements, which clearly show the temperature and magnetic field-dependent softening of the q-FMR mode [21,28]. At 80 K, where $TmFeO_3$ is primarily in $\Gamma_2$ phase in absence of magnetic field, the softening of the q-FMR mode occurs at a finite value of $\boldsymbol{H}_{DC}$, denoted as the critical field ($H_c$) ~1 T (Fig. 1a). At this critical field, the magnetic moments of the two Fe-sublattices reorient such a way that the net magnetic moment $\boldsymbol{m}$ aligns with the $\boldsymbol{H}_{DC}$ and the Néel vector remains orthogonal to $\boldsymbol{H}_{DC}$. Thus, $TmFeO_3$ undergoes a field-induced SRPT to the $\Gamma_4$ phase at higher fields. However, the frequency of the soft mode does not go to zero at the phase transition point ($H_{DC} = H_c$) likely due to magnetoelastic coupling with acoustic phonons [8,34,38,39]. We find the lowest frequency of the soft q-FM resonant mode ~9 GHz at $H_{DC} = H_c$. As the temperature is increased, the anisotropy energy for $\boldsymbol{n} \perp c$ ($\boldsymbol{m} \parallel c$) reduces [21,40]. This leads to lower critical field of SRPT. As a result, $H_c$ linearly reduces with higher temperature (Fig. 2g), consistent with theoretical prediction [28]. Notably, an additional softening of the q-FMR mode has been observed at zero external field at ~84 K which indicates the onset of the temperature-induced spontaneous $\Gamma_2 \to \Gamma_{24}$ phase transition. Further increase in temperature leads to shifting of the energy minimum towards $c$-axis. Thus $H_c$ further decreases with increasing temperature. This observation is also consistent with SQUID measurement as discussed in the supplementary information. At the onset of $\Gamma_{24} \to \Gamma_4$ phase transition, the weak ferromagnetic moment $\boldsymbol{m}$ spontaneously aligns along the $c$-axis (and $\boldsymbol{n} \parallel a$) [21] in the minimum energy configuration. The zero-field q-FMR mode is again a soft mode at the $\Gamma_{24} \to \Gamma_4$ phase transition [28] . This has been observed at ~94 K temperature in the magnetic resonance spectra where $H_c = 0$ (Fig. 2e).

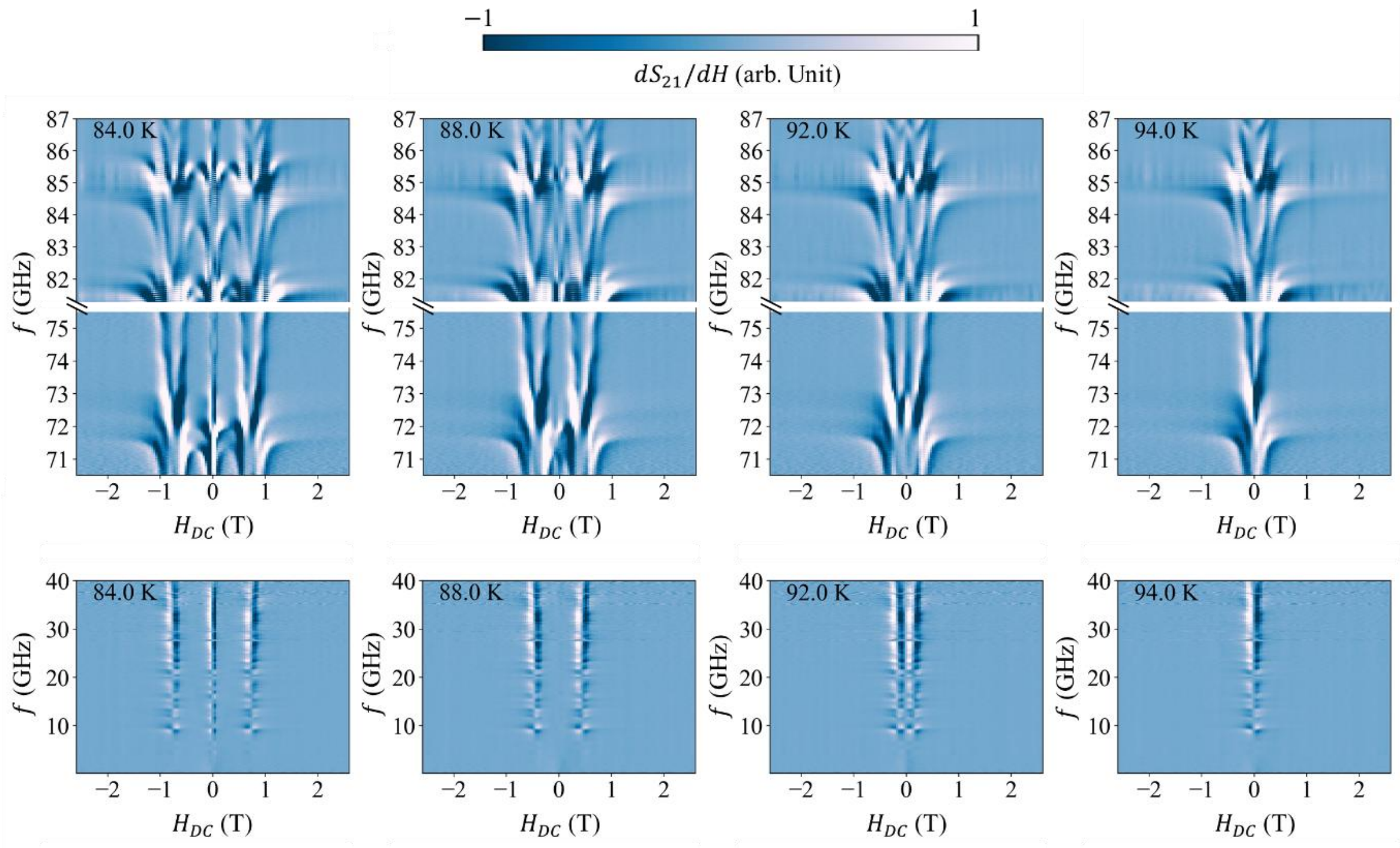

**Figure 3**. Full magnetic resonance spectra of $TmFeO_3$ for $H_{DC}$ along $c$-axis and the microwave magnetic field in the $ab$-plane.

In contrast to the low-frequency resonance spectra discussed above, the higher-frequency spectra of $TmFeO_3$ reveal several distinct modes as shown in Fig. 3. Despite the presence of multiple modes, these can still be attributed to q-FM precession of the sublattice magnetizations as the q-AFM resonant modes are expected to occur at much higher frequencies (around ~800 GHz) in this temperature range and are therefore inaccessible within our experimental frequency window [21,41]. These multiple q-FM modes in the 70.5–87.5 GHz band indicates excitation of several magnon modes with finite $\boldsymbol{k}$ vector which undergo softening like the q-FMR mode at the $\Gamma_2 \rightarrow \Gamma_{24}$ and $\Gamma_{24} \rightarrow \Gamma_4$ transition points (see Fig. 3). Notably, the q-FM magnon modes are most pronounced in the vicinity of the spin-reorientation transition. As the applied field $H_{\mathrm{DC}}$ increases further beyond the critical field $H_c$, the absorption amplitude of these modes decreases significantly, and their resonance frequencies display only weak magnetic-field dependence away from the transition region. This behavior reflects the strong magnetoelastic coupling between magnons and acoustic phonons which becomes dominant near the phase transition point [24,42].

From these results we find that the multiple high frequency q-FM modes exhibit similar field and temperature dependence and are shifted by ~0.5 GHz to 2 GHz with respect to each other. Notably, these additional modes are only observed in the intermediate phase $\Gamma_{24}$ (84 K $<$ T $<$ 94 K and $|\boldsymbol{H}_{DC}| < H_c$). In contrast, within the $\Gamma_2$ and $\Gamma_4$ phases, only one of the modes shows a strong field dependence (Fig. 3). The other modes are almost field-independent, and their amplitude decreases significantly when moving away from the transition points. The characteristic softening of the q-FM mode is consistent with previous observations [11,26,32] and predictions [21,28], as well as the behavior of the Goldstone mode (see also Supplementary Materials for simulations). Specifically, the $\Gamma_2 \rightarrow \Gamma_{24}$ and $\Gamma_4 \rightarrow \Gamma_{24}$ transitions are associated with broken rotational symmetry in the $ac$-plane which leads to the softening of the q-FMR mode. On the contrary, the excitation of multiple modes and their field dependence are surprising. In our experimental setup, we can only excite a uniform magnetic q-FMR mode. In the case of a single-domain sample, this mode is also decoupled from acoustic excitations. Therefore, to interpret this result, we first analyze the possible magnetic textures in the different phases. The magnetic symmetry of the $\Gamma_2$ and $\Gamma_4$ phases is orthorhombic which preserves the second order axes. This symmetry allows only two types of domains with opposite Néel vector orientations. Applying a static magnetic field perpendicular to the Néel vector ($\boldsymbol{H}_{DC} || a$ in phase $\Gamma_2$ and $\boldsymbol{H}_{DC} || c$ in phase $\Gamma_4$) removes the degeneracy between the domains, and the sample can be considered a single domain. In contrast, the symmetry of the intermediate $\Gamma_{24}$ phase is low enough to allow four types of domains with a generic orientation of the Néel vector in the $ac$ plane as schematically illustrated in Fig.4.

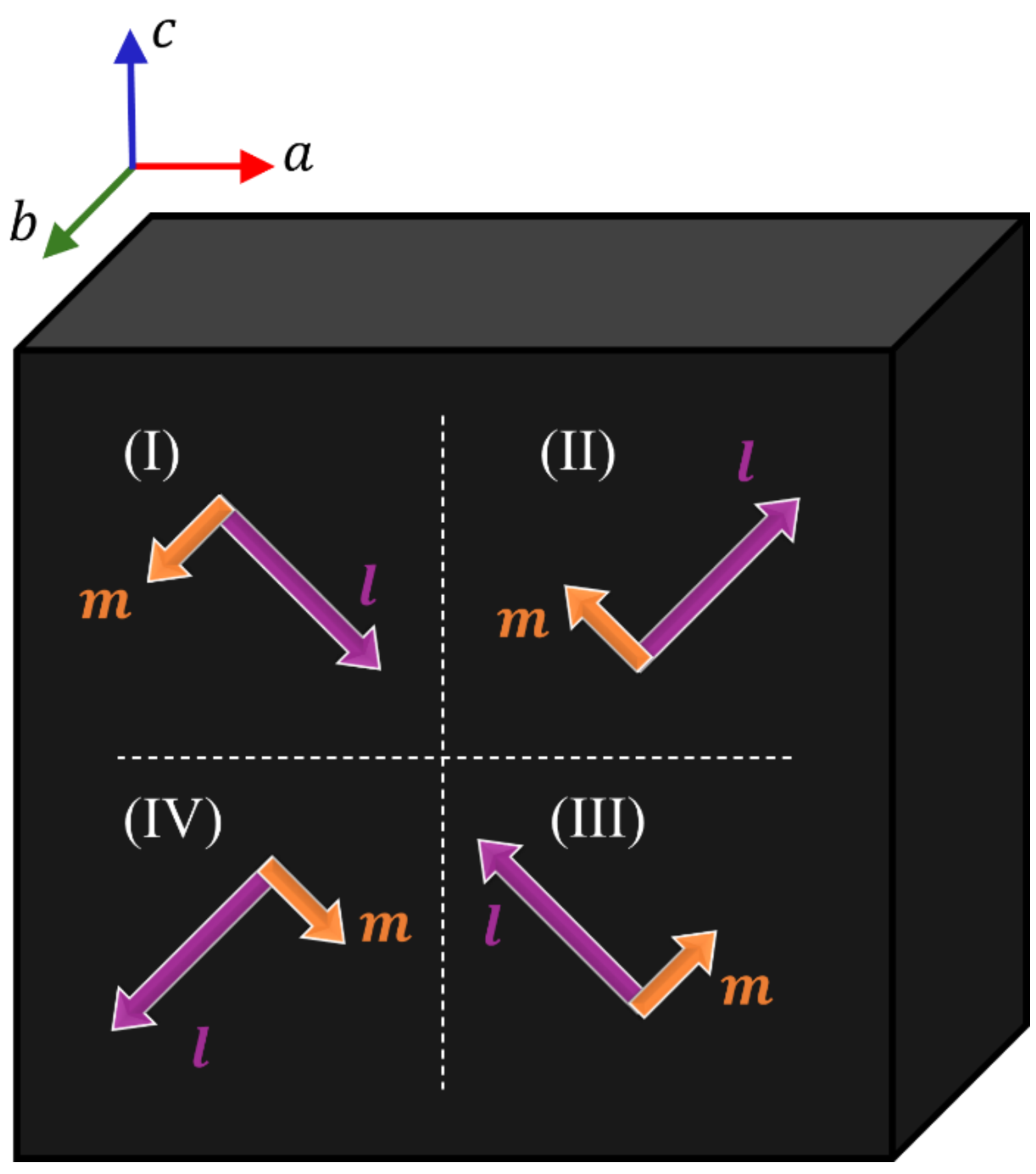

**Figure 4.** Orientation of $\boldsymbol{m}$ and $\boldsymbol{n}$ in four possible domain configurations in the $\Gamma_{24}$ phase.

When a magnetic field applied along the $c$ axis two of the four magnetic domains are energetically unfavorable, leaving domains whose magnetization components oppose the applied field (e.g., domains I and IV). The remaining non-180° domains (e.g., II and III) are energetically equivalent and possess opposite magnetization components along the $a$-axis. In our experimental geometry, these domains can be stabilized by the demagnetizing field along the $c$-direction, which is perpendicular to the sample surface. The domain structure in the intermediate $\Gamma_{24}$ phase was observed optically and it was reported that the period of the domain structure depends on the temperature [43]. Since the period of the domain structure depends on the energy of the domain walls, we can conclude that the period of the domain structure should also depend on the applied magnetic field in a similar fashion as it depends on the temperature. Consequently, unlike the $\Gamma_2$ and $\Gamma_4$ phases, the $\Gamma_{24}$ phase is magnetically inhomogeneous and can excite nonuniform modes in response to a uniform microwave excitation field. Additionally, the non-180° domains possess different spontaneous strains of the order of $5\times10^{-5}$. This means that the corresponding domain walls have magnetic and elastic components distributed at different spatial scales. Typically, the elastic component evolves on a much slower timescale than the magnetic component (the so-called frozen lattice approximation), resulting in pinned magnetic domain walls. In such a situation, an alternating magnetic field can excite not only the uniform mode but also the oscillation mode with a period equal to that of the domain structure. However, in the vicinity of a phase transition, the elastic and magnetic components evolve on the same timescale. The eigenmodes, in this case, include strongly hybridized magnetic and elastic components, consistent with the pronounced softening of the elastic modes as observed in the spin reorientation transition region [44]. The response of the periodic texture to a uniform alternating field includes a comb of frequencies. The observed 1

GHz difference between the modes corresponds to the estimated domain wall width (~0.5 μm). We assume that the domain walls where the Nèel vector rotates from the easy magnetic axis are the sources of the observed additional modes. Outside the transition region, the magnetic texture is homogeneous, and the additional hybridized modes are no longer excited.

We also cannot rule out the possibility of Suhl instability [45] occurring in the SRPT region. In particular, the magnetic energy in the $\Gamma_{24}$ phase contains a term that describes the coupling of a uniform q-FM mode with two transverse phonons that have opposite momentum and frequency half of the q-FMR mode. According to the classical model, this nonlinear interaction should induce parametric down-conversion of magnons into phonon modes. Due to strong magnetoelastic coupling, the efficiency of this process could be very high in the SRPT. Though we do not exclude this scenario, we consider this mechanism more speculative because we do not observe the expected threshold for this parametric process.

In summary, we have investigated the quasi-ferromagnetic magnon dynamics in $TmFeO_3$ single crystals across magnetic field and temperature-driven spin-reorientation phase transitions. Although the experimental geometry primarily excites the uniform q-FMR mode, additional q-FM spin-wave modes emerge in the intermediate $\Gamma_{24}$ phase. These modes likely originate from the finite domain structure and the associated strong hybridization between the q-FM magnons and acoustic phonons mediated by magnetoelastic coupling. Our findings highlight the important role of magnetoelastic interactions and magnetic textures in shaping the magnon spectrum of $TmFeO_3$ which may suggest new possibilities for actively tuning magnon modes in rare-earth orthoferrite systems.

**Acknowledgement**

This work has been supported by the German Research Foundation (DFG) under project No. 423441604. All authors from Mainz acknowledge support from SPIN + X (DFG SFB TRR 173 No. 268565370, projects A01, A12, and B02). M.K. acknowledges support from the Research Council of Norway through its Centers of Excellence funding scheme, project number 262633 "QuSpin". M.K. and E.G.-R. also acknowledge the support and funding from the project CRC TRR 288 - 422213477 Elasto-Q-Mat (project A12). This project has received funding from the European Union's Horizon Europe Program, Horizon 1.2 under the Marie Sklodowska Curie Actions (MSCA), Grant agreement No. 101119608 (TOPOCOM).

## References

[1] E. Bousquet and A. Cano, Non-collinear magnetism in multiferroic perovskites, J. Phys.: Condens. Matter **28**, 123001 (2016).

[2] A. Sasani, J. Íñiguez, and E. Bousquet, Origin of nonlinear magnetoelectric response in rare-earth orthoferrite perovskite oxides, Phys. Rev. B **105**, 064414 (2022).

[3] K. I. Tkachenko et al., Magnetic phase diagram of magnetocaloric$TmFeO_3$, Phys. Rev. B **113**, 054404 (2026).

[4] Y. Du, Z. X. Cheng, X. L. Wang, and S. X. Dou, Lanthanum doped multiferroic $DyFeO_3$: Structural and magnetic properties, J. Appl. Phys. **107**, 09D908 (2010).

[5] S. L. Gnatchenko, N. F. Kharchenko, P. P. Lebedev, K. Piotrowski, H. Szymczak, and R. Szymczak, Magneto-optical studies of H-T phase diagram for $DyFeO_3$ (H ‖ a), Journal of Magnetism and Magnetic Materials **81**, 125 (1989).

[6] F. J. Kahn, P. S. Pershan, and J. P. Remeika, Ultraviolet Magneto-Optical Properties of Single-Crystal Orthoferrites, Garnets, and Other Ferric Oxide Compounds, Phys. Rev. **186**, 891 (1969).

[7] D. Afanasiev, J. R. Hortensius, B. A. Ivanov, A. Sasani, E. Bousquet, Y. M. Blanter, R. V. Mikhaylovskiy, A. V. Kimel, and A. D. Caviglia, Ultrafast control of magnetic interactions via light-driven phonons, Nat. Mater. **20**, 607 (2021).

[8] V. D. Buchel'nikov, N. K. Dan'shin, L. T. Tsymbal, and V. G. Shavrov, Magnetoacoustics of rare-earth orthoferrites, Phys.-Usp. **39**, 547 (1996).

[9] V. G. Bar'yakhtar, B. A. Ivanov, and M. V. Chetkin, Dynamics of domain walls in weak ferromagnets, Sov. Phys. Usp. **28**, 563 (1985).

[10] V. D. Buchel'nikov, I. V. Bychkov, and V. G. Shavrov, Coupledoscillationsof iron rare-earth and elastic subsystems inorthoferrites with Kramers rare-earth ions, Zhurnal Eksperimental'noi i Teoreticheskoi Fiziki **101**, 1869 (1992).

[11] A. V. Kimel, C. D. Stanciu, P. A. Usachev, R. V. Pisarev, V. N. Gridnev, A. Kirilyuk, and Th. Rasing, Optical excitation of antiferromagnetic resonance in $TmFeO_3$, Phys. Rev. B **74**, 060403 (2006).

[12] A. V. Kimel, A. Kirilyuk, A. Tsvetkov, R. V. Pisarev, and T. Rasing, Laser-induced ultrafast spin reorientation in the antiferromagnet $TmFeO_3$, Nature **429**, 850 (2004).

[13] A. V. Kimel, B. A. Ivanov, R. V. Pisarev, P. A. Usachev, A. Kirilyuk, and T. Rasing, Inertia-driven spin switching in antiferromagnets, Nature Phys **5**, 727 (2009).

[14] T. T. Gareev, A. Sasani, D. I. Khusyainov, E. Bousquet, Z. V. Gareeva, A. V. Kimel, and D. Afanasiev, Optical Excitation of Coherent THz Dynamics of the Rare-Earth Lattice through Resonant Pumping of f-f Electronic Transition in a Complex Perovskite $DyFeO_3$, Phys. Rev. Lett. **133**, 246901 (2024).

[15] A. H. M. Reid, Th. Rasing, R. V. Pisarev, H. A. Dürr, and M. C. Hoffmann, Terahertz-driven magnetism dynamics in the orthoferrite $DyFeO_3$, Appl. Phys. Lett. **106**, 082403 (2015).

[16] L. Šmejkal, J. Sinova, and T. Jungwirth, Emerging Research Landscape of Altermagnetism, Phys. Rev. X **12**, 040501 (2022).

[17] A. V. Kimel, Th. Rasing, and B. A. Ivanov, Optical read-out and control of antiferromagnetic Néel vector in altermagnets and beyond, Journal of Magnetism and Magnetic Materials **598**, 172039 (2024).

[18] M. Naka, Y. Motome, and H. Seo, Altermagnetic perovskites, Npj Spintronics **3**, 1 (2025).

[19] E. Galindez-Ruales et al., Altermagnetic Magnon Transport in the d-Wave Altermagnet $LuFeO_3$, arXiv:2508.14569.

[20] X. Li, D. Kim, Y. Liu, and J. Kono, Terahertz spin dynamics in rare-earth orthoferrites, Photonics Insights **1**, R05 (2022).

[21] N. R. Vovk, E. V. Ezerskaya, and R. V. Mikhaylovskiy, Theory of terahertz-driven magnetic switching in rare-earth orthoferrites: The case of $TmFeO_3$, Phys. Rev. B **111**, (2025).

[22] S. Becker, A. Ross, R. Lebrun, L. Baldrati, S. Ding, F. Schreiber, F. Maccherozzi, D. Backes, M. Kläui, and G. Jakob, Electrical detection of the spin reorientation transition in antiferromagnetic $TmFeO_3$ thin films by spin Hall magnetoresistance, Phys. Rev. B **103**, 2 (2021).

[23] S. Schlauderer, C. Lange, S. Baierl, T. Ebnet, C. P. Schmid, D. C. Valovcin, A. K. Zvezdin, A. V. Kimel, R. V. Mikhaylovskiy, and R. Huber, Temporal and spectral fingerprints of ultrafast all-coherent spin switching, Nature **569**, 383 (2019).

[24] E. A. Turov and V. G. Shavrov, Broken symmetry and magnetoacoustic effects in ferro- and antiferromagnetics, Sov. Phy. Usp. **26**, 7 (1983)

[25] A. Bombik, B. Leśniewska J. Mayer, and A. W. Pacyna, Crystal structure of solid solutions $REFe_{1-x}(Al\ or\ Ga)_xO_3$ (RE=Tb, Er, Tm) and the correlation between superexchange interaction $Fe^{+3}$–$O^{-2}$–$Fe^{+3}$ linkage angles and Néel temperature, Journal of Magnetism and Magnetic Materials **257**, 206 (2003).

[26] S. Baierl, M. Hohenleutner, T. Kampfrath, A. K. Zvezdin, A. V. Kimel, R. Huber, and R. V. Mikhaylovskiy, Nonlinear spin control by terahertz-driven anisotropy fields, Nature Photon **10**, 715 (2016).

[27] R. C. LeCraw, R. Wolfe, E. M. Gyorgy, F. B. Hagedorn, J. C. Hensel, and J. P. Remeika, Microwave Absorption near the Reorientation Temperature in Rare Earth Orthoferrites, J. Appl. Phys. **39**, 1019 (1968).

[28] J. R. Shane, Resonance Frequencies of the Orthoferrites in the Spin Reorientation Region, Phys. Rev. Lett. **20**, 728 (1968).

[29] J. Zhang, M. Białek, A. Magrez, H. Yu, and J.-Ph. Ansermet, Antiferromagnetic resonance in TmFeO3 at high temperatures, Journal of Magnetism and Magnetic Materials **523**, 167562 (2021).

[30] J. A. Leakek and G. Shirane, THE MAGNETIC STRUCTURE OF THULIUM ORTHOFERRITE, $TmFeO_3$, Solid State Communications **6**, 1 (1968).

[31] R. A. Leenders, O. Y. Kovalenko, Y. Saito, N. R. Vovk, A. V. Kimel, and R. V. Mikhaylovskiy, THz-Driven Spin Dynamics in Orthoferrites with Kramers and Non-Kramers Rare-Earth Ions, Phys. Rev. Lett. **135**, 246703 (2025).

[32] K. Zhang, K. Xu, X. Liu, Z. Zhang, Z. Jin, X. Lin, B. Li, S. Cao, and G. Ma, Resolving the spin reorientation and crystal-field transitions in $TmFeO_3$ with terahertz transient, Sci Rep **6**, 23648 (2016).

[33] G. Gorodetsky and S. Shtrikman, Measurements of acoustic velocity and attenuation shifts at the spin-reorientation phase transition of $TmFeO_3$, Journal of Applied Physics **51**, 1127 (1980).

[34] R. M. Dubrovin et al., Spin and lattice dynamics at the spin-reorientation transitions in the rare-earth orthoferrite $Sm_{0.55}\ Tb_{0.45}\ FeO_3$, Phys. Rev. B **112**, 174419 (2025).

[35] E. Montoya, T. McKinnon, A. Zamani, E. Girt, and B. Heinrich, Broadband ferromagnetic resonance system and methods for ultrathin magnetic films, Journal of Magnetism and Magnetic Materials **356**, 12 (2014).

[36] H. Maier-Flaig, S. T. B. Goennenwein, R. Ohshima, M. Shiraishi, R. Gross, H. Huebl, and M. Weiler, Note: Derivative divide, a method for the analysis of broadband ferromagnetic resonance in the frequency domain, Review of Scientific Instruments **89**, 076101 (2018).

[37] I. Boventer, H. T. Simensen, A. Anane, M. Kläui, A. Brataas, and R. Lebrun, Room-Temperature Antiferromagnetic Resonance and Inverse Spin-Hall Voltage in Canted Antiferromagnets, Phys. Rev. Lett. **126**, (2021).

[38] V. I. Ozhogin and V. L. Preobrazhenskii, Nonlinear dynamics of coupled systems near magnetic phase transitions of the "order-order" type, Journal of Magnetism and Magnetic Materials **100**, 544 (1991).

[39] G. Gorodetsky, S. Shaft, and B. M. Wanklyn, Magnetoelastic properties of $TmFeO_3$ at the spin reorientation region, Phys. Rev. B **14**, 2051 (1976).

[40] H. Horner and C. M. Varma, Nature of Spin-Reorientation Transitions, Phys. Rev. Lett. **20**, 845 (1968).

[41] A. Hubert, Mikromagnetisch singuläre punkte in bubbles, Journal of Magnetism and Magnetic Materials **2**, 25 (1975).

[42] J. Zhang et al., Long decay length of magnon-polarons in $BiFeO_3/La_{0.67}Sr_{0.33}MnO_3$ heterostructures, Nat Commun **12**, 7258 (2021).

[43] A. I. Belyaeva, M. M. Kotlyarskii, and Yu. N. Stelmakhov, Visual investigation of spin reorientation in $TmFeO_3$, Phys. Stat. Sol. (a) **38**, K103 (1976).

[44] G. Gorodetsky and S. Shtrikman, Measurements of acoustic velocity and attenuation shifts at the spin-reorientation phase transition of $TmFeO_3$, Journal of Applied Physics **51**, 1127 (1980).

[45] V. I. Ozhogin and V. L. Preobrazhenskii, Nonlinear dynamics of coupled systems near magnetic phase transitions of the "order-order" type, Journal of Magnetism and Magnetic Materials **100**, 544 (1991).